\documentclass[useAMS,usenatbib]{mn2e}
\usepackage{graphicx}

\title[Photometry of V1062~Tau]
{Phototmetry of V1062~Tau: low states, short outbursts, and
period-switching}

\author[Y.~M.~Lipkin, E.~M.~Leibowitz and M.~Orio]
{Y.~M.~Lipkin$^1$\thanks{E-mail: yiftah@wise.tau.ac.il (YL);
    elia@wise.tau.ac.il (EML); orio@cow.physics.wisc.edu (MO)},
    E.~M.~Leibowitz$^1$\footnotemark[1] and
    M.~Orio$^{2,3}$\footnotemark[1]\\
 $^1$School of Physics and Astronomy and the Wise Observatory, Raymond
 and Beverly Sackler Faculty of Exact Sciences,\\ Tel-Aviv University,
 Tel Aviv, 69978, Israel;\\ 
 $^2$Istituto Nazionale di Astrofisica (INAF), Osservatorio
  Astronomico di Torino, Strada Osservatorio, 20, I-10025 Pino
  Torinese (TO), Italy\\
 $^3$Department of Astronomy, 475 N. Charter Street, University of
 Wisconsin, Madison, WI 53706, USA;} 

\begin{document}
\date{Accepted 2004 January 6. Received 2004 January 6; in original form 2003 July 31}

\pagerange{\pageref{firstpage}--\pageref{lastpage}} \pubyear{2004}

\maketitle

\label{firstpage}
\begin{abstract}
Time resolved photometry of the long-period intermediate polar
V1062~Tau confirmed the presence of the previously reported
orbital and spin periods,  and revealed the
presence of a third one, corresponding to the beat of the two.
While the orbital periodicity was present throughout our data, only
one of the shorter periods was detectable at any given time. 
 On a time-scale of $\sim$90 days, the short-period modulation in the
light curve of the star changed three times between the spin period and
the beat.
On longer time-scales, we report two outbursts of V1062~Tau (the
first to be recorded for this object) -- both of which were
probably short, low-amplitude ones ($\Delta{m}\sim1.2$~mag).
Our data also suggest a brief low state phase in 2002 January. 
Thus, this system joins two exclusive groups of
intermediate polars: those undergoing short outbursts, and those
having low states.

We propose that the alternations between the short periods that
modulate the light curve  were caused by changes in the accretion
mode, from disc-fed accretion, to disc-overflow accretion.
We further  suggest that these changes may have been triggered by
changes in mass-transfer rate, which were manifested by the
low-state/outburst activity of the system.
\end{abstract}


\section{Introduction}
Intermediate polars (IPs) are a subclass of Cataclysmic Variables
(CVs) having a white dwarf, the magnetic field of which is strong
enough to control (at least partially) the accretion flow from the
secondary star, but is not strong enough to synchronize the spin of
the primary star with the orbital revolution (for a review of IPs see
e.g. Patterson 1994; Warner 1995; Hellier 1995; Hellier 1996).

A prominent observational characteristic of IPs is multiple
 periodicities in the light curve, which may be modulated by the
spin and orbital periods, and a myriad of combinations of the two.

The intermediateness of the IPs, between the non-magnetic CVs and
the strongly-magnetic polars, makes this class a diverse one, which
includes stars with
a wide range of asynchronism and magnetic-field strengths.
The diversity of the class is further enhanced by the few accretion
modes found in IPs. 
While most systems are predominantly disc-fed, some accrete
predominantly in the disc-overflow mode (Hellier 1995) and one object
is a discless, stream-fed IP.

V1062~Tau (1 H 0459+248) was discovered with the {\it HEAO 1} Scanning
Modulation Collimator (A3 experiment) and the Large-Area Sky Survey
(A-1 experiment). 
A low-resolution spectrum with EXOSAT indicated that the source is a
cataclysmic variable (Remillard et al. 1994, hereafter R94).
The optical counterpart was identified as a UV-bright star, with
spectral features characteristic of a cataclysmic variable.
Time-resolved observations revealed two periodicities: 
a $P=9.952$~h period detected in optical photometry
was suggested to be the orbital period of the underlying binary system
($P_{orb} \equiv \Omega ^{-1}$; we note, however that confirmation by
radial velocity measurements is required).
Another periodicity, $P\sim62$~min, which was detected both in X-ray
and in the optical band, was naturally interpreted as the spin period
($P_{spin} \equiv \omega^{-1}$)
of the magnetic white dwarf in an intermediate polar system (R94). 
Further X-ray observations with ASCA and RXTE confirmed the suggested
classification of the system, and refined the spin period to
$P_{spin}=61.73\pm13$~min (Hellier, Beardmore \& Mukai 2002). 

Here we report on the results of a photometric campaign of V1062~Tau
in 2002 and 2003, during which the object exhibited unusual activity
both on short time scales and on long ones

\section{Observations and data reduction}
We conducted time-resolved photometry of V1062~Tau during 21 nights in
2001 Dec -- 2002 March. Additional 10 nights of time-resolved
photometry, and about three dozen snapshots were recorded in 2002
August -- 2003 March.
Most of the observations were conducted from the Wise Observatory in
Israel (WO), using the Tektronix 1K back-illuminated CCD mounted on
the 1-m telescope (a description of the telescope and the
instrumentation is given in Kaspi et al. 1995).

Our first run was a combined two-site operation with the Wisconsin
Indiana Yale NOAO (WIYN) 0.9-m telescope at the Kitt Peak National
Observatory (KPNO) in Arizona, USA.
The detector of the WIYN 0.9-m was the 2KSB CCD  (see the WIYN 0.9-m
internet page at http://www.noao.edu/0.9m/ for details about the
telescope and CCD).
The long temporal base-line between the two observatories allowed us
to achieve superior coverage in this run, as long as $\sim$20~h a
day. The nearly continuous coverage proved particularly useful for the
analysis of V1062~Tau because of its long orbital period and the
presence of additional periodicities in its light curve.

Photometry was conducted mostly through the $I$ filter.
A number of images were taken in $B$.
A few snapshots in $V$ and $R$ were also recorded.
A journal of our observations is given in appendix A.

Photometric measurements on the bias-subtracted and flat-field
corrected images were performed using the NOAO IRAF\footnote{IRAF
  (Image Reduction and Analysis Facility) is distributed by the
  National Optical Astronomy Observatories, which are operated by
  AURA, Inc., under cooperative agreement with the National Science
  Foundation.} {\sc{daophot}} package (Stetson 1987).
Instrumental magnitudes of V1062~Tau, as well as of a dozen or so
reference stars (depending on image quality), were measured in each
frame. 
We used the WO reduction program {\sc{daostat}} (Netzer et al. 1996)
to obtained a set of internally consistent magnitudes of the object
in each of the two data subsets acquired, by minimizing the scatter of the
reference stars over each subset.
We Cross-calibrated of the two subsets by setting the 
instrumental magnitudes of seven comparison stars to be the same in
both subsets.
Because of a possible difference in the spectral sensitivity of the
two instruments used, this procedure may introduce some second-order
systematic errors.
From measurements of the common reference stars, we estimate this
systematic effect to be less than $\sim0.03$~mag.
Finally, to transform our set of instrumental magnitudes to true
ones, we calibrated the reference stars using the photometric
standards of Landolt (1992).
The systematic error of the transformation is estimated to be less
than $\sim0.03$~mag.

\section{Results}
\subsection{Long-term variations}
\label{ObsLongTerm}
\begin{figure*}
\includegraphics[width=168mm]{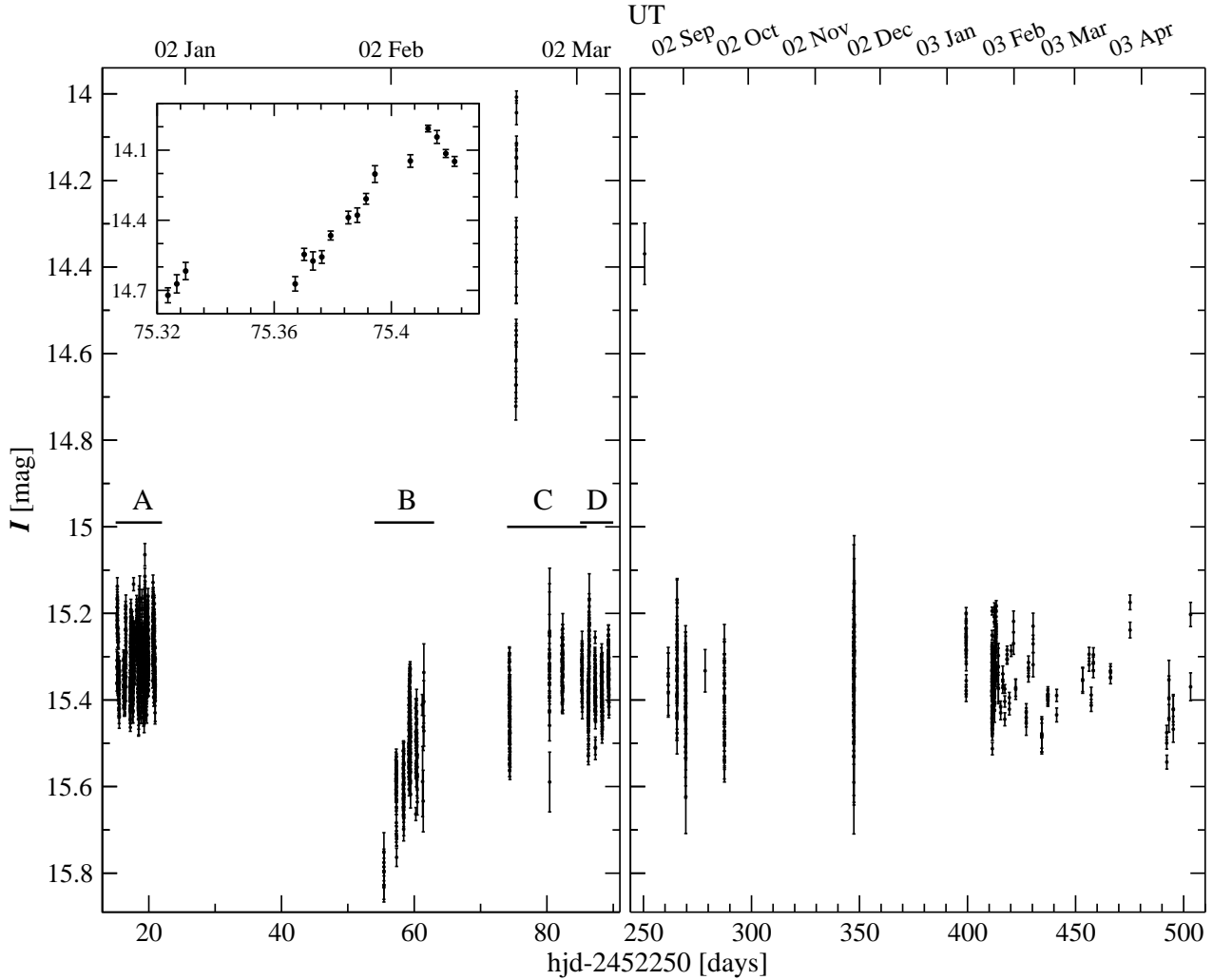}
\vskip 0.07in
\caption{Light curve of V1062~Tau in 2001-2003 ($I$-band). 
The left and right frames are the 2001-2002, and the 2002-2003
observations, respectively (note the difference in the scale of the  
$x$-axis between the two panels). The vertical lines are nightly 
observations, typically composed of a few tens of points. {\bf Inset:} a 
magnified view of the light curve of the outburst night of 2002 February 
19}
\label{figLC}
\end{figure*}

\begin{figure}
\includegraphics[width=84mm]{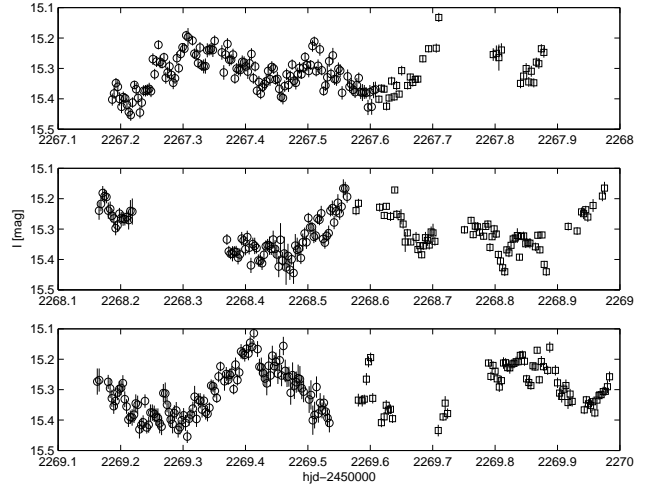}
\vskip 0.07in
\caption{A close-up view of three segments of the light curve from the 
  coordinated Wise-WIYN
  run. Circles are WO data, and squares are observations from
  WIYN.}\label{figNLC}
\end{figure}

The $I$-band light curve of V1062~Tau during 2001 December -- 2003
April is shown in figure~\ref{figLC}.
A detailed view of three segments of the light curve is given in
Fig.~\ref{figNLC}.
The light curve exhibits $\sim$0.2~mag modulations on a time-scale of
10~h, and weaker modulations on a time-scale of $\sim$1~h.
On longer time-scales, the object maintains constant brightness,
apart for three notable events.

On 2002 January 30 the star was $0.3$~mag fainter than in the
preceding observations, 34 days before.
Over the following $\sim$50 nights, the star gradually brightened, and
nearly returned to its brightness state of early 2002 January 
(Fig.~\ref{figLC}).
The rising trend of the light curve in February suggests a brief low state
of the star in 2002 January.

On 2002 February 19, V1062~Tau was in outburst ($\Delta{I}\ge1.16$~mag).
In subsequent observations five days later, the star was back in its
normal brightness level (Fig.~\ref{figLC}).
The outburst light curve is presented in figure~\ref{figLC} (inset).
The ending section of the outburst light curve is a 0.14~mag decline, which is
greater than the typical flickering of the light curve, and is not
phased with any of the system's periodicities.
This decline is therefore interpreted as beginning of the decline
from outburst maximum, making it a short
($\Delta{T_{outburst}}\approx1-2$~days), and weak outburst
($\Delta{I}=1.16$~mag).

Another outburst was detected in a snapshot taken on 2002  August 13,
wherein the star was 0.8~mag brighter than its normal state
(Fig.~\ref{figLC}).
The brighter state of the object was confirmed by a manual inspection
of the outburst frame.
The isolated observation does not allow us to set limits on the
amplitude or the duration of this outburst.

\subsection{Period analysis}
\label{ObsFourier}
\begin{figure*}
\includegraphics[width=168mm]{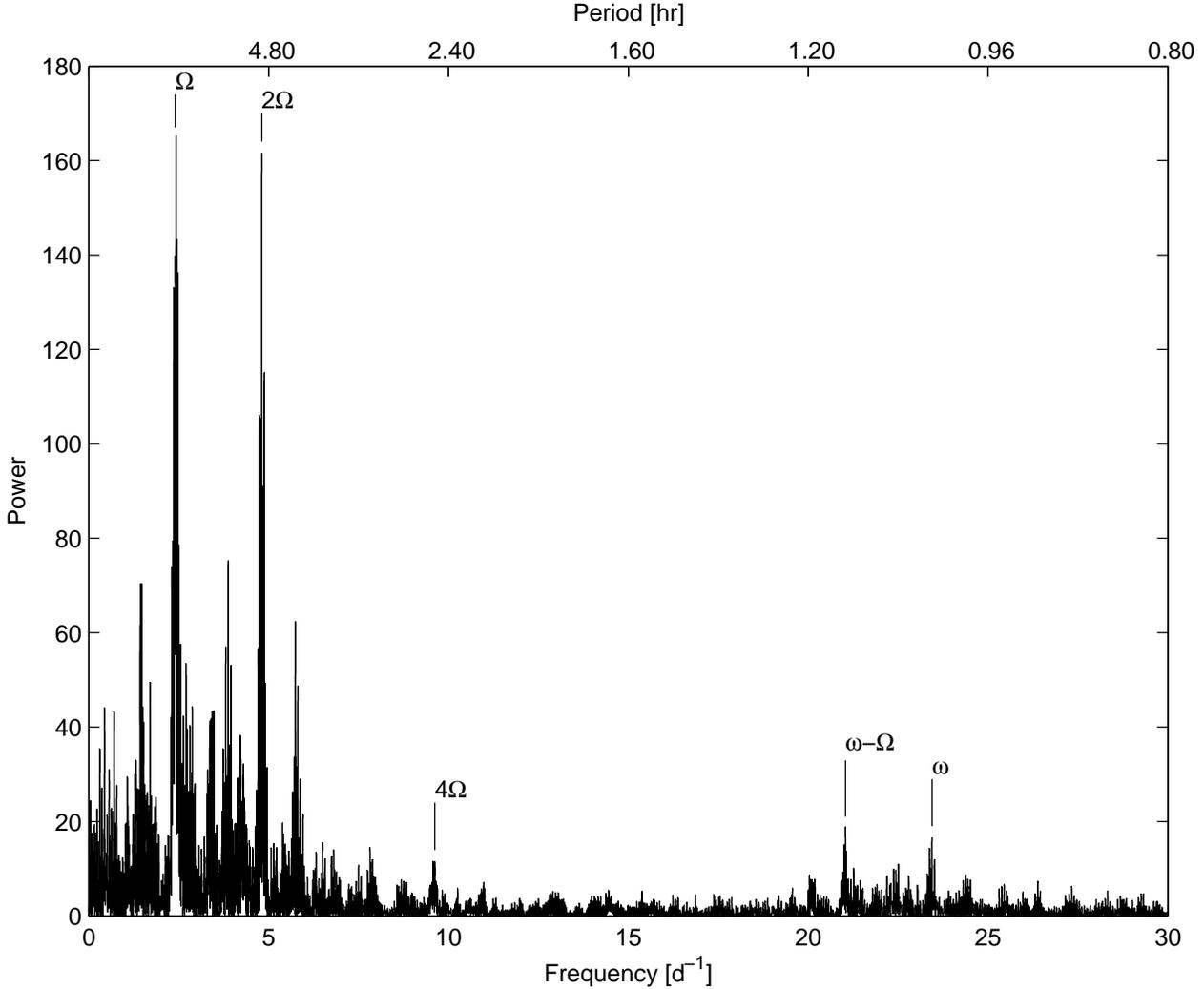}
\caption{Power Spectrum of the 2001-2002 $I$-band light curve of 
  V1062~Tau. The fundamental frequency and 1st and 3rd harmonics of
  the orbital period are $\Omega$, $2\Omega$, and $4\Omega$,
  respectively. The  fundamental frequencies of $P_{spin}$ and
  $P_{OSB}$ are marked by $\omega$ and $\omega-\Omega$, respectively.}
\label{figPS}
\end{figure*}

A power spectrum (PS, Scargle 1982) of the 2001-2002 light curve reveals
three independent periodicities in the data (Fig.~\ref{figPS}).
Two of the signals, the orbital period and the spin period were
previously reported by R94.
The third one, corresponding the first orbital sideband of the spin,
$P_{OSB}\equiv\left(P_{spin}^{-1}-P_{orb}^{-1}\right)^{-1}$ (synodic
period of the system) is reported here for the first time.

The lower end of the PS (Fig.~\ref{figPS}) is dominated by the
fundamental frequency of $P_{orb}$ at $9.9082\pm0.0006$~h, and by its  
first and third harmonics.
The high end of the PS features $P_{spin}$, at $61.43\pm0.05$~min,
and the orbital side band, $P_{OSB} = 68.45\pm0.09$~min.

The quoted periods were derived by finding a
minimum of $\chi^2$ for a simultaneous fit of three periodicities (see
Lipkin et al. 2001 for details of the analysis method).
The errors are 1-$\sigma$ confidence levels, estimated using the
bootstrap method (Efron \& Tibshirany 1993).
The periods we derived for $P_{orb}$ and $P_{spin}$ agree with former
measurements of the two (R94; Hellier et al. 2002).

\begin{figure*}
\includegraphics[width=168mm]{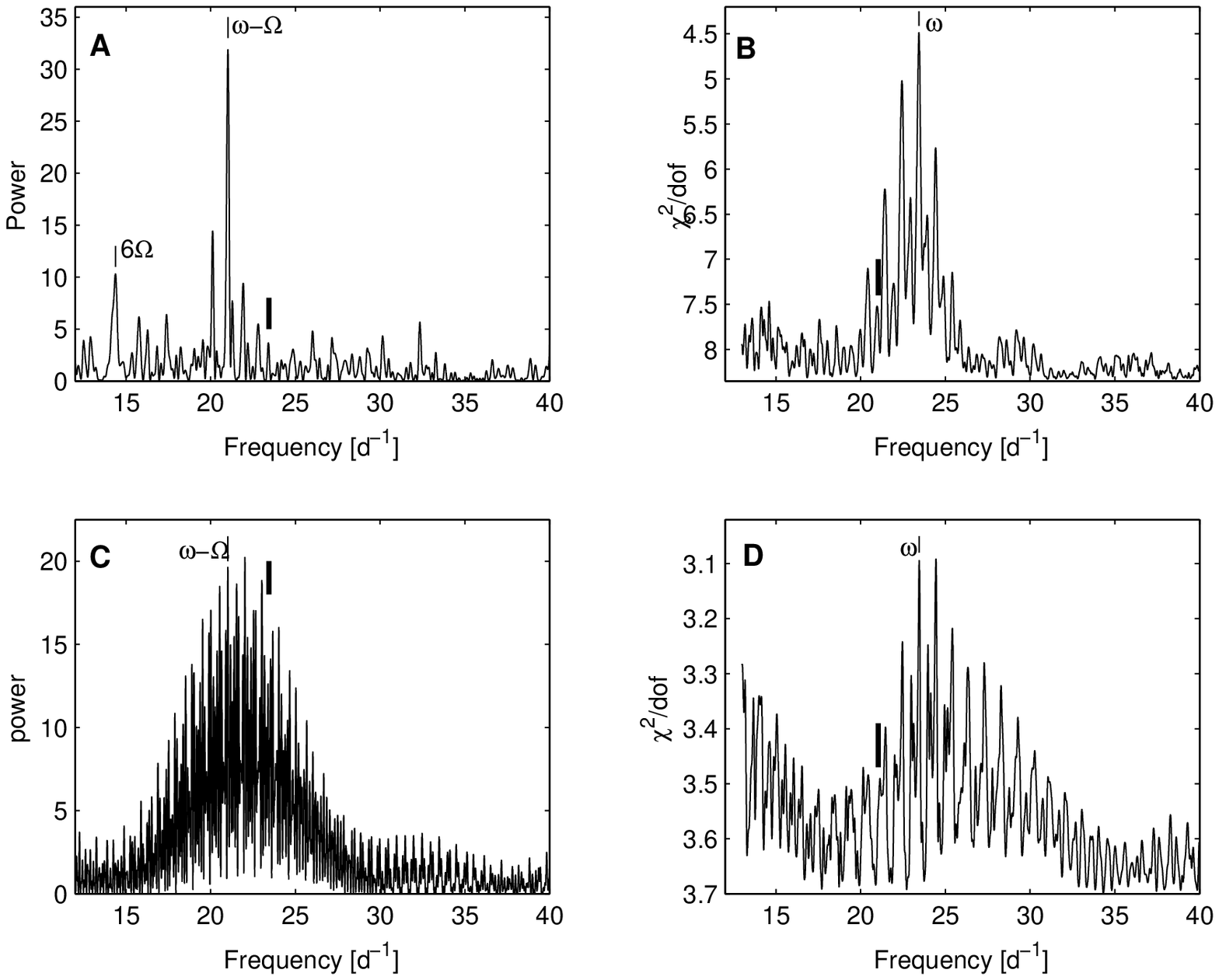}
\caption{{\bf Top left:} Power spectrum of dataset {\bf A}. The set
  was pre-whitened by subtracting the orbital signal.The presence of
  the orbital sideband in the data is  manifested by its fundamental
  frequency, and first harmonic. No significant signal is found at the
  spin period. The thick line marks the position of the spin period
  {\bf Top right:} $\chi^2$-periodigram of dataset {\bf B}. The orbital
  signal and a 2nd-power polynomial term were eliminated from the data
  in the calculation. The periodigram is dominated by an alias
  structure centred around the spin period. The orbital sideband is
  undetectable in the periodigram. The thick line marks to position of
  $P_{OSB}$.
  {\bf Bottom left:} Power spectrum of dataset {\bf C} (the night of
  the outburst was omitted from the analysis). The results
  are similar to those of set {\bf A}.
  {\bf Bottom right:} a $\chi^2$-periodigram of dataset {\bf D}. The
  results are similar to those of set {\bf B}.}
\label{figPSA}
\end{figure*}

Figure~\ref{figPSA} shows periodigrams of four subsets of our data
(the subsets are marked by the letters {\bf A}--{\bf D} in
Fig.~\ref{figLC}. Note a 2-night overlap between sets {\bf C} and {\bf
  D}).
The orbital signal is present in all four subsets.
However, the datasets yield different results in the high frequency
range.
In Sets {\bf A} and {\bf C}, $P_{OSB}$ is detected, but $P_{spin}$ is not
(the upper limits for the semi-amplitude of a signal at $P_{spin}$ in sets
{\bf A} and {\bf C} are 0.005, and 0.01~mag, respectively).
On the other hand, Sets {\bf B} and {\bf D} are dominated by the
signal of $P_{spin}$, and $P_{OSB}$ is not detected (with an upper
limit of 0.005, and 0.01~mag, respectively).
Analysis of a segment of our light curve, in which a transition from $P_{OSB}$
to $P_{spin}$ occurred (sets {\bf C} and {\bf D}, Fig.~\ref{figLC}),
revealed that the time-scale of the transition is two days.

Considering the changes of the short period in our data, we
re-examined frequency analysis of two nights of time-resolved
$I$-band photometry in presented in R94.
Using variance analysis, the authors deduced the existence of a single
short periodicity in the data, at $63.2\pm0.3$~min (Fig.~11 of R94).
Two shallower troughs, adjacent to the one selected by R94, were
probably interpreted by these authors as 1-day aliases. 
However, we find that these troughs, corresponding to $\sim$61.0, and 
$\sim$66.7~min,  are in better agreement with $P_{spin}$ and
$P_{OSB}$, respectively.
Another distinct trough, at $\sim$30.7~min, probably corresponds to
the first harmonic of $P_{spin}$.
We therefore conclude that during the 2 nights of R94's observations,
the star's light was probably modulated by both $P_{spin}$ and
$P_{OSB}$.
Naturally, reanalysis of the data of R94 is required in order to
verify this conclusion.

\subsection{Wave Forms}
\label{SecObsWF}
The 2001 December -- 2002 January light curve (set {\bf A}), folded over 
the
orbital period, and binned into 40 evenly-spaced bins is shown in
Fig.~\ref{figAmp} below (lower two frames; The signals of $P_{spin}$ and
$P_{OSB}$ were subtracted prior to the folding).
The orbital waveforms which we derived in different subsets agree with
the one shown in Fig.~\ref{figAmp} (although having greater
uncertainties because of the poorer sampling).

The asymmetric double-humped orbital waveform may be decomposed into a
fundamental component with a semi amplitude
$\Delta{I}=0.063\pm0.001$~mag, and a first harmonic with
$\Delta{I}=0.047\pm0.001$~mag,  which is lagging by
$\Delta{\phi}=0.099\pm0.005$~cycle. 
As noted by R94, ellipsoidal variations of the secondary are likely to
be the source of the first harmonic.
The leading fundamental component may arise from a few different
sources, such as aspect variation of a bright spot at the outer edge
of the accretion disc, or reflection effect, where light from
the bright spot is reflected by the secondary.
A detailed analysis of the relative phases between the different
components of the orbital modulation would allow us to locate the
various sources taking part in the orbital modulation (e.g. the hot
spot) in the binary reference frame.

\begin{figure}
\includegraphics[width=84mm]{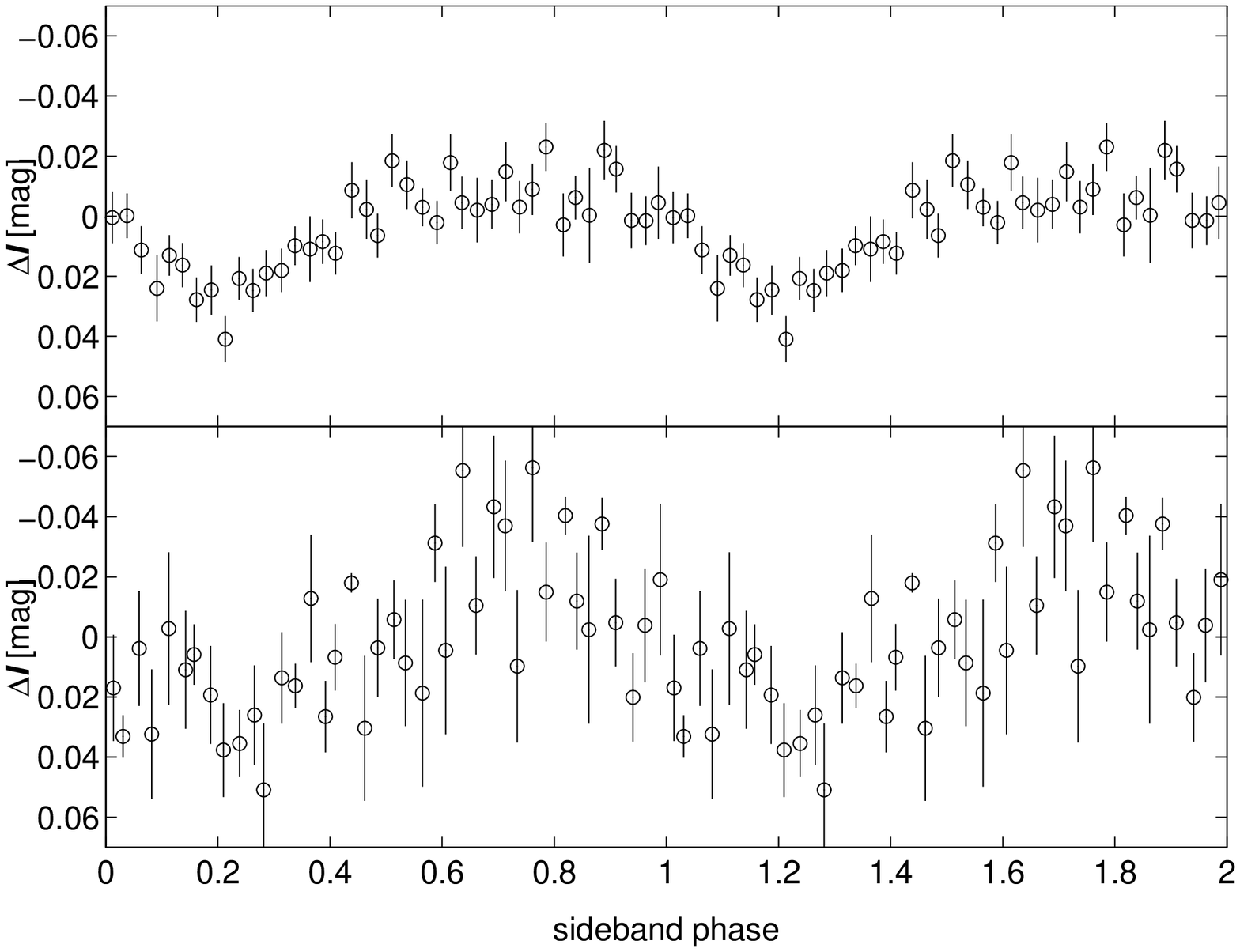}
\caption{{\bf Upper frame:} subset {\bf A}, folded over $P_{OSB}$, and
  binned into 40 evenly-spaced bins. The orbital signal was removed
  prior to the folding {\bf Lower frame:} same as above, for set {\bf
    C}. The same (arbitrarily determined) phase zero was used for both sets.}
\label{figSBWF}
\end{figure}

The waveform of $P_{OSB}$ (Fig.~\ref{figSBWF}) maintained the phase
and a similar symmetric shape in both subsets in which it was
detected.
However, the amplitude  of the signal in the first set was half the
amplitude in the second one (0.05 and 0.1~mag in sets {\bf A} and {\bf
  C}, respectively). 

\begin{figure}
\includegraphics[width=84mm]{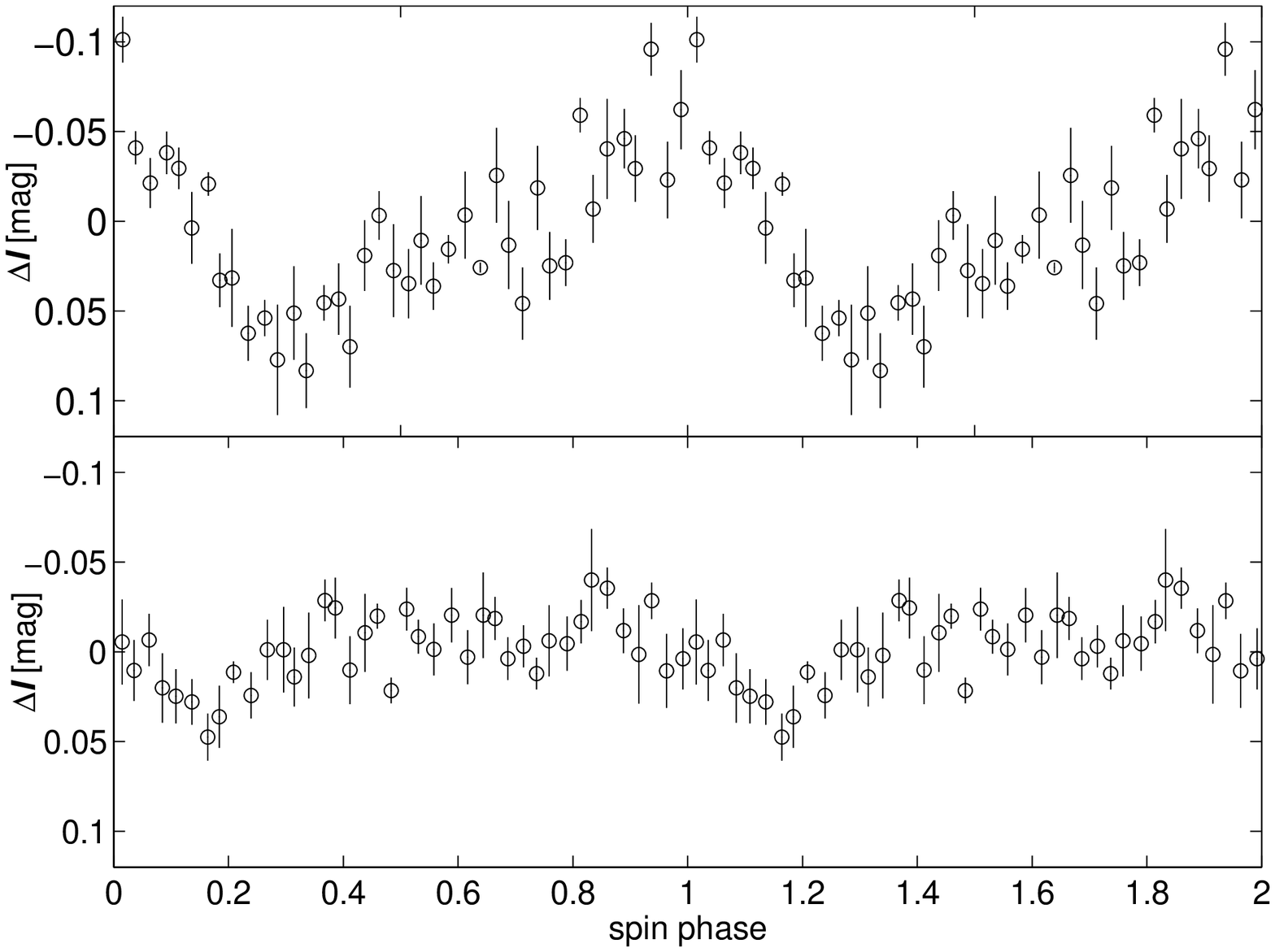}
\caption{{\bf Upper frame:} subset {\bf B}, folded over $P_{spin}$, and
  binned into 40 evenly-spaced bins. The orbital signal and a secular
  trend were removed prior to the folding {\bf Lower frame:} same as
  above, for set {\bf D}. The same (arbitrarily determined) phase zero
  was used for both sets.}
\label{figSpinWF}
\end{figure}

The amplitude of the spin also changed considerably between set {\bf
  B} and set {\bf D}, decreasing by half in the latter ($\Delta{I}=$ 0.18, and
0.09~mag, respectively, Fig.~\ref{figSpinWF}).
The spin waveform also changed between the two sets, both in phase and
shape, however the significance of these differences is questionable. 

\subsection{Variation of the short signals over the orbital cycle}
\label{SecObsCorr}
\begin{figure*}
\includegraphics[width=168mm]{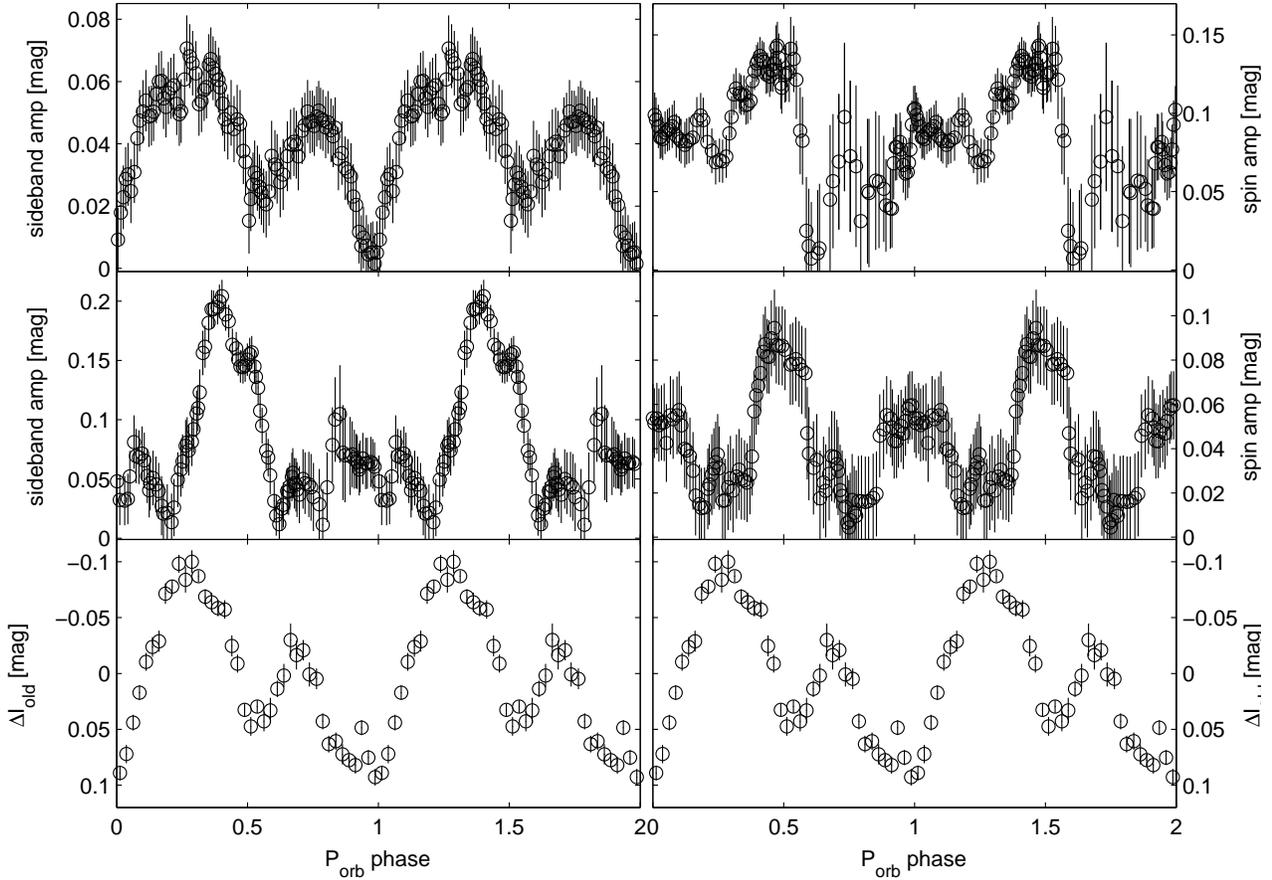}
\caption{The amplitudes of the spin and sideband modulations in
  different orbital phases. The orbital waveform is shown in the
  bottom frames, for comparison.
{\bf Upper left:} the amplitude of the sideband
  signal, in set {\bf A} (the phase interval used is
  $\Delta{\phi}=0.15$, see description in the text).{\bf Centre left:}
  same as above, in set {\bf C} ($\Delta{\phi}$ used is 0.22). {\bf
  Upper right:} the spin amplitude in set {\bf C}
  ($\Delta{\phi}=0.24$). {\bf Centre Right:} same as above, for set
  {\bf D} ($\Delta{\phi}=0.20$).}
\label{figAmp}
\end{figure*}

Measurements of the spin and sideband amplitudes over different
orbital phases are shown in Fig.~\ref{figAmp}.
The upper- and middle-left panels show the measured amplitudes of
$P_{OSB}$ in sets {\bf A} and {\bf C}, respectively.
The measurements of the spin amplitude in set {\bf B} and set {\bf D}
are shown in the upper- and middle-right panels. 
The bottom panels show the orbital waveform, for reference use.

The amplitudes at different orbital phases $\phi_0$ were derived by
fitting a sine wave to a subset of data points, located within a small
phase interval $\Delta{\phi}$ about $\phi_0$.
The error bars in figure~\ref{figAmp} are the calculated
1-$\sigma$ errors of the fit.

The amplitude of $P_{OSB}$ varied strongly over the orbital cycle,
acquiring a maximum value twice every orbital cycle.
Indeed, the $P_{OSB}$ amplitude is strongly correlated with the
changing brightness of the star over the orbital cycle. 
In set {\bf A} the correlation coefficient is $\rho=0.8\pm0.1$, and the 
phase lag is $\Delta\phi=0.005$, whereas in set {\bf C}, $\rho=0.8\pm0.1$, and
$\Delta\phi=0.15$. 
Note the different phase lags, as well as the different amplitudes of
the two sets (Fig.~\ref{figAmp}). 

Similarly, measurements of the spin  over different orbital
phases revealed that the spin amplitude also varied strongly over the
orbital phase.
For the spin signal we obtained $\rho=0.5\pm0.1$, and $\Delta\phi=0.11$ in 
set {\bf B}, and $\rho=0.64\pm0.09$, $\Delta\phi=0.24$ in set {\bf D}.

\section{Discussion}
\subsection{Long-term variations}
Our data suggest a low state of V1062~Tau in 2002 January, only the
ascent of which was observed (Sec.~\ref{ObsLongTerm}).
In the (sparse) published material of V1062~Tau we find three
additional occasions in which V1062~Tau was possibly observed in
low-state.
The object was observed by R94 in five different occasions between
1988 February 22 -- 1990 December 11.
In one (1988 February 22), the object was 3-4 times fainter.
Munari, Zwitter \& Bragaglia (1997) present a spectrum of the star
from 1995 November 4, which yielded an integrated $V$ magnitude of $16.6$
-- about a magnitude fainter than the star's normal brightness level
which was reported by R94.
Finally, we  note that R94 reported that V1062~Tau is UV bright
(indeed this was the criterion by which they identified the optical
counterpart of the  X-ray source). 
A blank spectrum which was obtained after four hours of integration
with IUE on 1994 November 3 (Szkody \& Silber 1996) may imply a low
state of the star during this epoch.

Low state are observed in all well-studied polars, as well as in many
non-magnetic CVs (see Warner 1999 for a recent review of low states
in CVs).
However only two IPs (both with orbital periods $P_{orb}\la$4~h) have
been observed in low states.
Low states appear to come in two types. 'slightly low' states are
brightness reduction with an amplitude $\la$1~mag, usually having
recurrence time-scales which are characteristic for each object
(typically 5--10 yr). 
This type is observed only in non-magnetic CVs.
In contrast 'very low' states are more extreme drops in brightness,
with no characteristic recurrence time-scales, occurring both in
magnetic and non-magnetic CVs.

Although commonly attributed to reduced mass-transfer from the
secondary star, the underlying mechanism of low states is still poorly
understood.
'Very low' states have been proposed to be the result of migration of
star-spots through L1, reducing the  Roche-lobe overfill of the
secondary at L1 (Livio \& Pringle 1994). 
'Slightly low' states may be the effect of magnetic cycles in the
secondary star (see e.g. Warner 1988).

Garnavich and Szkody(1988) found in a sample of magnetic and
non-magnetic CVs that for systems with orbital periods
$P_{orb}\ga200$~min, the magnitude difference between low- to
high-state increases with $P_{orb}$.
For V1062~Tau, low states are expected to be $\ga$6~mag fainter than
the normal state. 
That out of three recorded low-states in V1062~Tau, two 
were observed only $\sim$1~mag fainter than its mean brightness may
suggest that low states of V1062~Tau do not follow the expected trend.
Alternatively, the time V1062~Tau spends in low states may be short
relative to the ascension/descension times.

There are a number of IPs which undergo dwarf-nova like outbursts.
However, two systems exhibit shorter, low amplitude outbursts, which
are suggested to be the result of increased mass transfer from the
secondary (Hellier et al. 2000; Hellier, Mukai \& Beardmore 1997): 
V1223~Sgr (only one observed outburst, van Amerongen \& van Paradijs,
1989), and TV~Col (a few short outbursts, e.g. Hellier \& Buckley,
1993).
Another IP, EX~Hya, shows outbursts on intermediate time scales
($\sim2-3$~d; Hellier et al. 2000)).
It is yet unclear whether these events are the result of
disc-instabilities, or mass-transfer enhancements (Hellier et al. 2000).

The 2002 February 19 outburst of V1062~Tau, which was only  partially
recorded (Sec.~\ref{ObsLongTerm}),  should probably be classified as
a short burst. 
The 0.14~mag decline at ending section of the outburst light curve
probably marked the start of the decline from maximum brightness
(Fig.~\ref{figLC}, inset).
Additionally, the upper limit for the duration of the 2002 February
outburst is $T\le4.9$~d -- less that half the expected duration for a
dwarf nova-like outburst of a $P_{orb}\sim$10~h CV (Warner, 1995).
The outburst of 2002 August 13, which was recorded only by a single
snapshot, is also consistent with a short-type burst
(Sec.~\ref{ObsLongTerm}).  
Therefore, V1062~Tau is probably a new member in the exclusive group of
IPs having short outbursts.

Finally, the different characteristics of the outburst of 2002 February 
19, and of the 2002 Jan low state suggest that the two are caused by
different instability mechanisms in the secondary star.
The superposition of the short outburst on the recovering branch of
the low state present a remarkable interplay between the two
different instabilities.

\subsection{Short-term signals}
During 2001 December -- 2002 March V1062~Tau oscillated with either the 
spin, or the orbital sideband period, switching three times between
the two. The transition time-scale between these two modes was
$\sim$2~d. The characteristic time scale for maintaining a
period was $\sim$5--$\sim$25~d (Sec.~\ref{ObsFourier}).
The amplitude of both periods was correlated with the orbital phase,
with different phase lags in the different subsets.

Modulations at the spin period are the signature of a disc-fed IP
system.
Possible sources of such variations are aspect variation of the
accretion zone (e.g. Hellier et al. 1987),
or reprocessing of X-ray emission in axisymmetric regions of the
disc, either the inner edge of the magnetically-truncated disc, and/or
the outer rim of a concave disc (Warner 1986).
Sideband modulations may also occur in disc-fed systems, arising from 
reprocessing of X-ray light in fixed locations in the binary reference
frame: the secondary star (Patterson \& Price 1981), and/or the bright
spot bulge on the outer rim of the accretion disc (Hassall et
al. 1981).
In predominantly stream-fed IPs, optical sideband modulations may
also arise from the accretion stream flipping between the magnetic
poles twice every synodic cycle. 

The detection of both $P_{spin}$ and $P_{OSB}$ in the $I$-band light
curve of V1062~Tau, and the absence of both in $V$
(see Sec.~\ref{ObsFourier} and Fig.~11 of R94), suggest that the
source of the two signals is more likely to be reflection rather than
emission from the hotter accretion region.
The strong variation of the amplitude of both signals over the
orbital cycle (sec.~\ref{SecObsCorr}, Fig.~\ref{figAmp}) also supports
reflection as the source of the modulations.
Had these variations emanated from the accretion region of the
CV system, strong variability of the amplitudes of the short
periodicities would require the obscuration of most of this inner
region during some part of the orbital cycle.
However the orbital wave form is hardly an eclipsing one.
If the accretion region is not obscured, any flux
modulations it may exhibit should not be affected by the orbital cycle.
Thus, measured in magnitude units, changes in the amplitude of such
modulations over orbital cycle should only reflect the varying
relative contribution of the fixed signal the to the changing total
brightness of the system.
This is hardly the case in V1062~Tau (Fig.~\ref{figAmp}).
Finally, the changes in the characteristics of the short signals
between different data sets (Fig.~\ref{figAmp}) would be difficult to
account for if these variation are emission from the accretion
region.

On the other hand, a simple reflection model also seems insufficient
to account for the properties of the short-term modulations. In
particular, the changes between the different data sets and the form
of the amplitude dependence on the orbital phase introduce similar
difficulties when ascribing the short periodicities to reflection.

The transitions between $P_{spin}$ and $P_{OSB}$ imply changes
in the accretion mode and/or in the geometry of the system.
A similar effect had been observed in the X-ray light of TX~Col -- a
long period IP ($P_{orb}=5.72$~h), in which both the spin and the
sideband are observable.
The relative intensities of the two signals were found to vary
by an order of magnitude over a typical time-scale of $\la$1~month
(Buckley \& Tuohy 1989; Norton et al. 1997; Wheatley 1999).
Changes in the relative power of the sideband signal were also
observed in FO~Aqr (Beardmore et al. 1998), AO~Psc, V1223~Sgr and
BG~CMi (e.g. Hellier 1998).
Such changes are usually attributed to variable amounts of
disc- and disc-overflow accretion.
The cause of such transitions is not clear.
X-ray observations of TX~Col suggest that periods of enhanced
disc-overflow are related to decreased $\dot{M}$ (Wheatley 1999).
However changes in the accretion mode may also be driven by various
mechanisms in the binary system  which are not related to changes in
$\dot{M}$ (see, e.g., Norton et al. 1997).

The period switching which we observed in the optical light of
V1062~Tau may also reflect changes in the accretion mode, between
predominantly disc-accretion and predominantly disc-overflow accretion.
The exclusive appearance of the two signals may imply that during
periods of predominantly disc-overflow accretion, disc-accretion
became negligible or ceased completely, making the disc a
'nonaccretion-disc' during these periods (see King \& Lasota 1991).
The average brightness of V1062~Tau was not considerably different
during periods in which the system's light  was modulated by different
a short-period modulation (see Fig.~\ref{figLC};
particularly, the average brightness of set {\bf C} and that of set
{\bf D} are indistinguishable).
However, we note the close coincidence of period switching events with
the low-state/outburst activity of the star:
one transition from $P_{OSB}$ to $P_{spin}$ coincided with the
apparent low state of 2002 Jan, and another occurred shortly
after the short outburst of 2002 February 19.

Period switching may also be caused as a result of changes in
scale-height of the accretion disc, 
if both short signals emanate from reprocessed light (e.g. the beat is
reflection from the  bright spot bulge and/or from the secondary
star, and the spin  is reflection from the outer rim of a concave
disc).
Orbital sideband modulations would occur during periods when the
accretion disc is thin.
Conversely, when the disc rim thickens considerably  (e.g. following
changes in mass-transfer rate from the secondary), it may obscure the
secondary star, and the axi-symmetric outer rim of the disc would
become the main source of reflection, varying at the spin period.
It would be hard, however to reconcile such a change in the accretion
disc with the stable average magnitude of V1062~Tau over the periods of
changing short modulation.

\section{Summary}
\begin{enumerate}
\item We report two outbursts of V1062~Tau, the first to be reported for
this object. One, and possibly both bursts are  short, low amplitude
ones. 
V1062~Tau is thus a member of the exclusive group of  IPs harboring
short bursts.

\item Our observations and previously published data imply four
different low state events of V1062~Tau.
Only two other IPs, both with much shorter orbital periods, are known
to have low states.

\item During 2002 December -- 2003 March, the light of V1062~Tau was 
modulated
by the orbital period and either of two short periods: $P_{spin}$, and
$P_{OSB}$.
during our observations the star switched three times between the two
short period. Our data suggests that the time-scale of transition is
$\sim$2~d.

\item We suggest that a likely mechanism for the transition between the
two short periods is temporal changes in the accretion mode, changing
between disc-fed, and disc-overflow accretion.
\end{enumerate}

\section*{Acknowledgments}
This work is supported by grants from the Israel Science Foundation.
YML is grateful to the Dan-David prize foundation for financial support.

\appendix
\section[]{Log of observations}

\begin{tabular}{@{}ccccccc@{}}
UT&Time of Start&Run Time&\multicolumn{4}{c}{Points per Filter}\\
   Date  &(HJD-2452200) & (h)& I & B & V & R\\

20011221 &  65.18 &  9.85 & 55   & 55 &  &  \\
20011222 &  66.17 & 10.01 & 43   & 43 &  &  \\
20011223 &  67.18 & 10.12 & 140  &    &  &  \\
{\bf 20011223} & {\bf 67.61} & {\bf 6.34} &  {\bf 39} & \multicolumn{3}{c}{{\bf(WIYN0.9m)}}\\
20011224 &  68.16 &  9.54 & 87   &    &  &  \\
{\bf 20011224} & {\bf 68.58} & {\bf 9.56} & {\bf 74}  & \multicolumn{3}{c}{{\bf(WIYN0.9m)}}\\
20011225 &  69.16 &  9.99 & 135  &    &  &  \\
{\bf 20011225} & {\bf 69.81} & {\bf 9.66} & {\bf 74}  & \multicolumn{3}{c}{{\bf(WIYN0.9m)}}\\
{\bf 20011226} & {\bf 70.58 }& {\bf 9.67} & {\bf 97}  & \multicolumn{3}{c}{{\bf(WIYN0.9m)}}\\
20020130 &  104.43 &  0.36  & 6    &    &  &  \\
20020201 &  107.25 &  2.29  & 33   &    &  &  \\
20020202 &  108.35 &  2.13  & 31   &    &  &  \\
20020203 &  109.19 &  6.25  & 89   &    &  &  \\
20020204 &  110.22 &  6.23  & 28   & 28 &  &  \\
20020205 &  111.21 &  5.91  & 11   & 8  &  &  \\
20020218 &  124.31 &  3.17  & 45   &    &  &  \\
20020219 &  125.32 &  2.35  & 23   &    &  &  \\
20020224 &  130.36 &  1.44  & 21   &    &  &  \\
20020226 &  132.27 &  3.29  & 46   &    &  &  \\
20020301 &  135.24 &  2.11  & 23   & 2 & 2 &  \\
20020302 &  136.20 &  4.50  & 61   &   &   &  \\
20020303 &  137.23 &  2.63  & 38   &   &   &  \\
20020304 &  138.20 &  3.93  & 56   &   &   &  \\
20020305 &  139.22 &  3.14  & 45   &   &   &  \\
20020318 &  152.26 &  0.16  & 1    & 1 & 1 &  \\
20020813 &  300.56 &  0.07  & 1    &   &   &  \\
20020824 &  311.51 &  0.69  & 9    &   &   &  \\
20020828 &  315.52 &  1.99  & 29   &   &   &  \\
20020901 &  319.49 &  2.84  & 32   &   &   &  \\
20020910 &  328.55 &  0.08  & 1    & 1 &   &  \\
20020919 &  337.44 &  1.77  & 26   &   &   &  \\
20020926 &  344.57 &  0.07  & 1    &   &   &  \\
20021118 &  397.27 &  7.32  & 84   &   &   &  \\
20030109 &  449.34 &  1.05  & 16   &   &   &  \\
20030120 &  460.27 &  0.15  & 3    &   &   &  \\
20030121 &  461.31 &  4.83  & 60   & 1 & 1 &  \\
20030122 &  462.41 &  2.67  & 33   & 1 & 1 &  \\
20030123 &  463.23 &  3.63  & 15   & 1 & 1 &  \\
20030124 &  464.24 &  1.19  & 6    & 1 & 3 &  \\
20030125 &  465.33 &  0.07  & 2    &   &   &  \\
20030126 &  466.31 &  0.14  & 3    &   &   &  \\
20030127 &  467.26 &  0.14  & 3    &   &   &  \\
20030128 &  468.32 &  0.14  & 3    &   &   &  \\
20030129 &  469.34 &  0.08  & 2    &   &   &  \\
20030130 &  470.19 &  0.33  & 1    & 1 & 1 & 1\\
20030131 &  471.32 &  0.08  & 2    &   &   &  \\
20030201 &  472.30 &  0.08  & 2    &   &   &  \\
20030206 &  477.19 &  0.16  & 3    &   &   &  \\
20030207 &  478.21 &  0.18  & 3    &   &   &  \\
20030209 &  480.34 &  0.23  & 3    &   &   &  \\
20030213 &  484.36 &  0.14  & 3    &   &   &  \\
20030216 &  487.18 &  0.15  & 3    &   &   &  \\
20030220 &  491.24 &  0.08  & 2    &   &   &  \\
20030304 &  503.28 &  0.08  & 2    &   &   &  \\
20030307 &  506.26 &  0.08  & 2    &   &   &  \\
20030308 &  507.22 &  0.08  & 2    &   &   &  \\
20030309 &  508.24 &  0.14  & 3    &   &   &  \\
20030317 &  516.20 &  0.14  & 6    &   &   &  \\
20030326 &  525.24 &  0.15  & 3    &   &   &  \\
20030412 &  542.22 &  0.17  & 3    &   &   &  \\
20030413 &  543.21 &  0.16  & 3    &   &   &  \\
20030415 &  545.21 &  0.15  & 3    &   &   &  \\
20030423 &  553.23 &  0.51  & 2    & 1 & 1 & 1\\
20030424 &  554.21 &  0.15  & 3    &   &   &  \\
\end{tabular}

\label{lastpage}
\end{document}